\documentclass[a4 paper,12pt]{article}
\normalbaselines
\catcode `\@=11
\@addtoreset{equation}{section}

\def\section{\@startsection {section}{1}{\z@}{-3.5ex plus -1ex minus
     -.2ex}{2.3ex plus .2ex}{\normalsize\bf}}
\def\subsection{\@startsection{subsection}{2}{\z@}{-3.25ex plus -1ex
minus
 -.2ex}{1.5ex plus .2ex}{\normalsize\bf}}

\def\thebibliography#1{\section*{References}
\list
  {[\arabic{enumi}]}{\settowidth\labelwidth{[#1]}\leftmargin\labelwidth
  \advance\leftmargin\labelsep
  \usecounter{enumi}}
  \def\newblock{\hskip .11em plus .33em minus -.07em}
  \sloppy
  \sfcode`\.=1000\relax}
 
\catcode `\@=12
\newcommand{\mathbb}[1]{{\bf #1}}
\begin{document}

\vspace*{2.5cm}
\begin{center}

{\bf ALGEBRODYNAMICAL APPROACH IN FIELD THEORY:\\
 BISINGULAR SOLUTION AND ITS MODIFICATIONS}
\vspace{1.3cm}\\
\medskip

{\bf  V. V. Kassandrov, J. A. Rizcalla}
 \vspace{0.3cm}\\
 Department of General Physics, Russian People's Friendship University,\\
 Moscow, Ordjonikidze 3, 117419, E-mail: vkassan@mx.pfu.edu.ru.	\\ 
 \end{center}

\vspace*{0.5cm}

{\small
\noindent
The criterion of differentiability of functions of quaternion variable is used as the basis of some algebraic
field theory. Its necessary consequences are free Maxwell and Yang-Mills equations. The differentiability equations may
be integrated in twistor variables and are reduced to algebraic ones. In the article we present bisingular solution 
and its topological modifications. Related EM-fields appear to be just the 
well-known Born solution and its modifications with singular structure of 
ring-like and toroidal topology. General problems of the algebrodynamical 
approach are discussed.
}

\section{\hspace{-4mm}.\hspace{2mm}INTRODUCTION}
\label{int}
Physics turns back to its basic principles.
Nowadays, it claims not only to {\it describe} reality, though in a mathematically elegant and well extrapolated
way. The question concerns the {\it essence} of physical laws themselves, and the existence of a general physical
principle governing all natural processes.

It becomes evident that such a principle may be expressed in a purely abstract mathematical language only and 
manifests itself in some {\it programme}, {\it "Code of Universe"} predetermining the structure and evolution of the World.
From this point of view, {\bf the laws of nature should be worked out "on paper" not in the laboratory.}
The true language of nature might be drastically different from that being used in modern physics, 
and one should then reject the correspondence principle (in the now adopted narrow sense).

The feasibility of such a paradigm certainly finds difficulty in the unreadiness of most physicists to abandon the 
habitual "Newton-Galilean" methods and eclectic medley of classical, quantum and geometric notions, and to make 
an attempt to build up physics all over again, "on the sheet of paper". More wonder is that for the present mathematics 
itself isn't able to offer some really unique structure, which could naturally generate the {\it heterogeneity} and
{\it hierarchy}, so inherent to the World.

Perhaps, only {\it fractal mappings} may easily reproduce the most simple of these properties, giving birth to
a whole world "from Nothing". Another possibility on which we will concentrate in this paper is related 
to a {\it hyper-holomorphic} structure, which is a generalization of the Cauchy-Riemann
criteria in complex calculus, to some exclusive algebras of 
{\it quaternionic} type. Based solely on this structure it was found possible to build
a non-Lagrangian field theory, where the {\it noncommutativity} of these algebras leads
to {\it nonlinearity} of the generalized  Cauchy-Riemann criteria making it
feasible to include physical interactions in this approach\footnote
{Contrary to many methods where quaternionic calculus was
based on direct {\it linear} generalizations of the Cauchy-Riemann criteria.}.

Preliminary results and some variants of this approach, which was called 
{\it algebrodynamics}, were given in~\cite{AD}. In the most interesting and studied 
case the equations defining a holomorphic function on the algebra of 
biquaternions $\mathbb{B}$, are reduced to

\begin{equation}\label{eq1}
d\eta = \Phi * dX * \eta ,
\end{equation}
where $\eta$ - 2-spinor, $\Phi$ - 2$\times$2 matrix gauge field (for details see Sec.~\ref{gse}).

It turned out that system (\ref{eq1}) is closely related to the fundamental equations of 
physics~--~{\it vacuum} Maxwell, Yang-Mills and Einstein equations, 
as well as to d'Alembert and eikonal equations \cite{AD}-\cite{K34}. Precisely, {\bf 
for every solution of system (\ref{eq1}) these equations are satisfied 
identically}\footnote{Vacuum Einstein equations are satisfied only for static 
metrics (see Sec.~\ref{fnd}).}, this itself being an astonishing fact!

It was found also that the class of singular solutions of the vacuum 
equations themselves is wide enough to contain solutions with not only
point-like and ring-like singularities, but also {\it toroidal, helix-like} and 
even more complex structure of singularities. This allows us  in the framework of our model 
to consider {\bf particles as field singularities}~\cite{GC,K34}, and try to relate their transmutations 
to "perestroikas" of singularities precisely in the sense of the catastrophe theory~\cite{Arn}.   

The most important, however, is the nonlinear and overdetermined 
system (\ref{eq1}) itself, consistently defining both the 2-spinor $\eta(x)$ and the gauge 
field $\Phi(x)$. For solutions of system (\ref{eq1}) {\bf the superposition principle is
violated}, so it serves as a {\it "filter"} selecting those solutions of the field equations 
which are compatible with it and determining their evolution in time, i.e. interactions and 
transmutations of singularities-particles.

On the other hand and contrary to customary field theory, the overdetermined 
system (\ref{eq1}) (8 equations for 6 functions) does not allow to fix even the initial 
field distribution in all $\mathbb{R}^3$. In full accordance with 
Einstein's {\it super-causality concept}~\cite[P. 762]{Ein} this ensures on the classical level "quantization rules" 
for some chracteristics of these particles-singularities (such as the electric charge)~\hbox{\cite{AD,GC,Klu,CCF}})!

The most difficult problem is the construction of {\it multisingular} solutions and the establishment of 
the general form of the equation of motion for these singularities as a consequence of the field equations (\ref{eq1}). 
After a brief survey of properties of system (\ref{eq1}) (Sec.~\ref{gse}) and its integration
in twistor variables (Sec.~\ref{twt}) we obtain in an algebraic way the formerly found 
fundamental unisingular solution (Sec.~\ref{fnd}), present the exact bisingular
solution of system (\ref{eq1}) and discuss its modifications (Sec.~\ref{bsm}). In Sec.~\ref{con} we 
conclude our results and point out some general questions of the algebrodynamical approach.

\section{\hspace{-4mm}.\hspace{2mm}MAIN PROPERTIES OF THE GENERATING SYSTEM}
\label{gse}
The system of equations (\ref{eq1}) (hereafter called the generating system of equations, GSE) is
a particular form of the differentiability criteria of a function of biquaternionic
variable. A variant of noncommutative calculus corresponding to this case was
set forth in details in \cite{AD,GC,Klu} (references to former works are to be found 
therein). The multiplication ($*$) in formula (\ref{eq1}) could be regarded as a matrix 
one, and the field $\Phi(x)$ - as a complex $2\times2$ matrix, in accordance with
the well-known representation of the algebra of biquaternions $\mathbb{B}$.

Under the Lorentz transformations
\begin{equation}\label{eq2a}
X \Rightarrow A^+ * X * A , ~~~ \det A =1 ,
\end{equation}
GSE conserves its form provided the quantities $\eta(x)$ and $\bar{\Phi}(x)$
(a matrix adjoint to $\Phi(x)$) behave as a 2-spinor and a 4-vector respectively:
\begin{equation}\label{eq2b}
\eta \Rightarrow \bar{A}*\eta , ~~~\bar{\Phi} \Rightarrow A^+*\bar{\Phi}*A .
\end{equation}

The GSE is also covariant under the {\it "weak"} gauge transformations, i.e. 
transformations of the form
\begin{equation}\label{eq3}
\eta \Rightarrow \lambda(\eta) \eta , ~~~\Phi_\mu \Rightarrow \Phi_\mu + \frac{1}{2}
\partial_\mu \lambda(\eta),
\end{equation}
with a gauge parameter $\lambda(\eta_1,\eta_2)$ depending on the coordinates only 
implicitly through components of the affected solution \cite{GC,CCF}.

Hence, the quantity $\Phi(x)$ may be considered as a 4-potential of a complex
abelian gauge field, with dynamical restrictions which follow from commutation relations 
and have the form of complex {\it self-duality conditions} \cite{AD,GC}
\begin{equation}\label{eq4}
\vec E + i \vec H = 0
\end{equation}
for the $\mathbb{C}$\,-valued fields corresponding to $\Phi(x)$. In view of
relations (\ref{eq4}) Maxwell equations are satisfied identically and separately for real and 
imaginary parts of field strengths $\vec{E}$ and $\vec{H}$. That is why the $\mathbb{C}$\,-valued
field $\Phi(x)$ actually describes a {\it pair} of dually-conjugate electromagnetic fields.

Moreover, the field $\Phi(x)$ admits a supplementary interpretation \cite{GC,K34}.
Indeed, the left $\mathbb{B}$\,-connection
\begin{equation}\label{eq5}
\Gamma(x) = \Phi(x) * dX \equiv \Gamma^0(x) + \Gamma^a(x)\sigma_a
\end{equation}
in a natural way defines a strength of some complex gauge field
\begin{equation}\label{eq6}
F(x) = d\Gamma(x) - \Gamma(x) \wedge \Gamma(x).
\end{equation}

The part $\Gamma^0(x)$=$\Phi_\mu(x)dx^\mu$ corresponds to the above treated
electromagnetic field. At the same time the traceless part $\Gamma^a(x)$ of 
connection (\ref{eq5}), expressed in terms of $\Phi_\mu(x)$, due to relations
(\ref{eq4}) {\bf satisfies free Yang-Mills equations} \cite{GC}. Observe that the 
electromagnetic $F^0_{[\mu,\nu]}$ and Yang-Mills $F^a_{[\mu,\nu]}$ strengths are related by
\begin{equation}\label{eq7}
F^a_{[\mu,\nu]} F^a_{[\mu,\nu]} = (F^0_{[\mu,\nu]})^2 
\end{equation}
(summation is over the isotopic index $a=1,2,3$ only), which is different from
the relation accepted in customary field theory.

Other aspects of GSE (\ref{eq1}), including a geometrical interpretation of connection
(\ref{eq5}) as a {\it Weyl-Cartan connection} of a special type, can be found in \cite{GC,Klu,CCF}.

\section{\hspace{-4mm}.\hspace{2mm}TWISTOR VARIABLES AND INTEGRATION OF GSE}
\label{twt}
Equations (\ref{eq1}) may be written in a 2-spinor form
\begin{equation}\label{eq8}
\nabla^{AA^\prime} \eta^{B^\prime} = \Phi^{B^\prime A} \eta^{A^\prime}, ~~~~A,A^\prime,B^\prime= 1,2 ,
\end{equation}
where $\nabla^{AA^\prime}$ stands for the derivative with respect to the
spinor coordinates $X_{AA^\prime}$. Mutilplying by an orthogonal spinor 
$\eta_{A^\prime}$=$\varepsilon_{A^\prime C^\prime}\eta^{C^\prime}$ we exclude the 
potentials $\Phi^{B^\prime A}$ and come to the {\bf generalized nonlinear 
Cauchy-Riemann equations} for the spinor $\eta(x)$ components:
\begin{equation}\label{eq9}
\eta_{A^\prime} \nabla^{AA^\prime} \eta^{B^\prime} = 0 .
\end{equation}

The solution of (\ref{eq9}) may be expressed in terms of twistor variables
\begin{equation}\label{eq10}
\tau_A = X_{AA^\prime} \eta^{A^\prime}
\end{equation}
in an implicit with respect to $\eta(x)$ form
\begin{equation}\label{eq11}
\Psi^{(C)}(\eta,\tau) = 0, ~~~C = 1,2, 
\end{equation}
where $\Psi^{(1)},\Psi^{(2)}$ are two arbitrary holomorphic functions of the four complex variables
\hbox{$\eta^{1^\prime},\eta^{2^\prime},\tau_{1},\tau_{2}$.}

Solving equations (\ref{eq11}) for the spinor components 
$\eta^{1^\prime},\eta^{2^\prime}$ and substituting the result in (\ref{eq8}) we can 
determine the potentials $\Phi^{C^\prime A}(x)$ and, consequently, the 
electromagnetic and Yang-Mills fields corresponding to the solution of 
system (\ref{eq9}).

Chiefly we are interested in field's {\it singularities}: their topological
structure, electric charge and evolution in time. Differentiating (\ref{eq11}), it is 
easy to show that for the {\it caustics}, where the fields blow up, we have
\begin{equation}\label{eq12}
\det \|\frac{d\Psi^{(C)}}{d\eta^{B^\prime}}\| = 0 .
\end{equation}

For subsequent analysis, let us note that the solution of (\ref{eq11}) is simplified by 
fixing the gauge to $\eta^{1^\prime}$=1. Under this gauge the first of 
equations (\ref{eq11}) becomes trivial: $\Psi^{(1)}$=$\eta^{1^\prime}-1$=0 and the
remaining spinor component $\eta^{2^\prime} \equiv~G(x)$ is determined 
algebraically from the second relation of (\ref{eq11})
\begin{equation}\label{eq13}
\Psi^{(2)}(\eta^{2^\prime},\tau_1,\tau_2) \equiv \Psi(G,~ wG+u,~ vG+\bar w) = 0,
\end{equation}
where the following notation was adopted for spinor coordinates:~$u,v=X_{11^\prime},X_{22^\prime}=t \pm z$,
~~$w,\bar{w}=X_{12^\prime},X_{21^\prime}=x \pm iy$.
The caustics equations reduce then to a simpler form  
\begin{equation}\label{eq14}
\frac{d\Psi}{dG} = 0.
\end{equation}

Actually, equations (\ref{eq9}) in the above gauge reduce to the well-known relations,
defining a {\it null shear-free geodesical congruence}, while their solution (\ref{eq13})
is given by the Kerr theorem \cite[Chapter 7]{Pen}. {\bf Singularities of the
corresponding metric and its curvature} \cite{K34,KDS,KW} {\bf are determined by the caustic 
equation (\ref{eq14})} \cite{Bur,Bur1}.

Hence the singularities of the EM, Yang-Mills and gravitational fields associated with solutions of GSE (\ref{eq1}) are
determined by the same relation (\ref{eq14}) and therefore coincide. That's why the here
accepted interpretation of {\bf particles as common singularities of these fields} seems quite natural.  

\section{\hspace{-4mm}.\hspace{2mm}FUNDAMENTAL UNISINGULAR SOLUTION OF GSE}
\label{fnd}
A choice of a linear function $\Psi(G,\tau_1,\tau_2)$ leads to a trivial solution
$G(x)$ with identically zero fields. {\it Static axisymmetric} solutions are given
by functions (\ref{eq13}) of the form 
$$
\Psi = G\tau_1- \tau_2 - cG = 0, ~~~c = Const,
$$
or equivalentely
\begin{equation}\label{eq15}
G(wG+u) - 2cG - (vG+\bar w) = 0.
\end{equation}

The solution of the second degree equation (\ref{eq15}) is 
\begin{equation}\label{eq16}
G(x) = \frac{\bar w}{(z-c)\pm\sqrt{(z-c)^2 + \rho^2}} ,
\end{equation}
where $z=(u-v)/2,~\rho^2=w\bar w=x^2+y^2$. For real-valued $c$ which can be
eliminated by a proper translation, solution (\ref{eq16}) defines a 
{\it stereographic projection} $S^2 \to \mathbb{C}$, and electromagnetic field of
the Coulomb type
\begin{equation}\label{eq17}
E_r = \frac{q}{r^2},~~q = \pm 1; ~~~E_{\theta} = E_{\varphi} = 0,
\end{equation}
with a {\bf fixed value of the electric charge} (so called {\it "algebraic charge quantization"}
analogous to (\ref{eq17}) monopole form
\begin{equation}\label{eq18}
H_r = \frac{i\mu}{r^2}, ~~\mu = q = \pm 1; ~~H_{\theta} = H_{\varphi} = 0.
\end{equation}

In case of imaginary values of the constant $c=ia,~a\in \mathbb{R}$ the 
field singularity has a {\it ring-like} structure with radius $r=|a|$. For $r\gg |a|$
the field has a multipole structure with Coulomb-monopole main terms. This
enables one to estimate the quadrupole moment of particles corresponding to solution (\ref{eq16})~\cite{K34}.

The Riemannian metric for solution (\ref{eq16}) is of Schawrzshild and Kerr type
for real and imaginary values of the constant $c$ respectively \cite{KDS}. In \cite{KW} it was shown 
that {\bf the only static empty space metric of the Kerr-Schild type whose 
singularities (as well as EM field singularities in our approach) are confined to 
a bounded region are those corresponding to solutions (\ref{eq16})}. It is remarkable that fields corresponding to 
(\ref{eq16}) automatically satisfy vacuum Einstein equations as well as electrovacuum Maxwell-Einstein system of equations. Indeed, the 
energy-momentum tensor $T_{\mu\nu}$ for the complex self-dual EM-fields, satisfying (\ref{eq4}), vanishes identically! 
On the other hand, Maxwell equations in the considered spaces are identical to those in flat space~\cite{KDS}.

\section{\hspace{-4mm}.\hspace{2mm}BISINGULAR SOLUTION AND ITS MODIFICATIONS}
\label{bsm}
Let us consider a time-dependent axisymmetric solution, generated by a function
$$
\Psi = \tau_1 \tau_2 + b^2G = 0,~~~ b=Const \in \mathbb{C}\,.
$$
Again, we have a second degree equation with the following solution:
\begin{equation}\label{eq19}
G = \frac{-2u \bar w}{\sigma^2 + \rho^2 + b^2 \pm \sqrt{\Delta}},~~~~\Delta \equiv 
(\sigma^2 + \rho^2 + b^2)^2 - 4\sigma^2 \rho^2,
\end{equation}
where $\sigma^2 = uv =t^2-z^2 $.

The caustic for this solution is determined by a zero discriminant $\Delta=0$, which may be transformed to
the following {\it equation for singularities' evolution}:
\begin{equation}\label{eq20}
\Delta = (t^2 - z^2 - \rho^2 + b^2)^2 + 4 b^2 \rho^2 = 0.
\end{equation}
In case of real constant $b$, as a solution of (\ref{eq20}) we have
\begin{equation}\label{eq21}
\rho = 0, ~~~ z = \pm \sqrt{t^2 + b^2},
\end{equation}
so that the field has two point-like singularities moving with 
constant acceleration (i.e. performing mirror {\it hyperbolic} motion) along $\pm Z$ axis (Fig.1a,b,c). 
Calculating the electric flux through closed surfaces surrounding each 
singularity we find that they {\bf possess electric charges equal in 
magnitude to the charge of fundamental solution and opposite to each 
other.}
\vskip5mm
\newcommand{\jss}[1]{\normalsize #1}
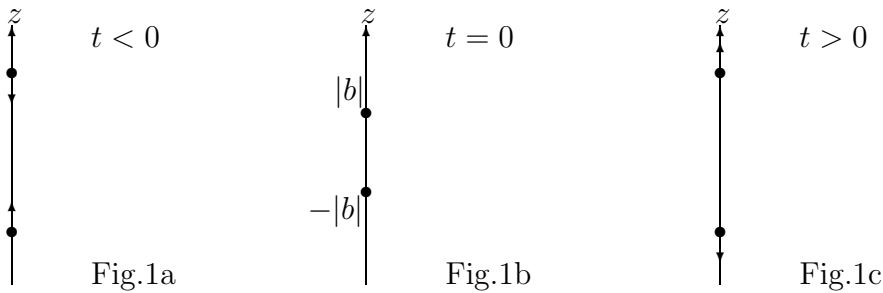
\begin{picture}(100,100)
\put(50,0){\vector(0,1){98}}
\put(48,99){\jss{$z$}}
\put(80,90){\jss{$t<0$}}
\put(80,0){\jss{Fig.1a}}
\put(50,80){\circle*{4}}
\put(50,20){\circle*{4}}
\put(50,80){\vector(0,-1){12}}
\put(50,20){\vector(0,1){12}}
\end{picture}
\begin{picture}(100,100)(-30,0)
\put(50,0){\vector(0,1){98}}
\put(48,99){\jss{$z$}}
\put(80,90){\jss{$t=0$}}
\put(80,0){\jss{Fig.1b}}
\put(50,65){\circle*{4}}
\put(50,35){\circle*{4}}
\put(38,70){\jss $|b|$}
\put(28,25){\jss $-|b|$}
\end{picture}
\begin{picture}(100,100)(-60,0)
\put(50,0){\vector(0,1){98}}
\put(48,99){\jss $z$}
\put(80,90){\jss $t>0$}
\put(80,0){\jss{Fig.1c}}
\put(50,80){\circle*{4}}
\put(50,20){\circle*{4}}
\put(50,80){\vector(0,1){12}}
\put(50,20){\vector(0,-1){12}}
\end{picture}
\vskip5mm
The electromagnetic field (precisely its real part $\Re \vec E,\Re \vec H$) generated by  
this solution has the form of the well-known {\it Born solution} for a field of uniformly 
accelerated charge (\cite[P. 136]{Pal})
\begin{equation}\label{eq22}
 E_\rho = \mp \frac{8b^2\rho z}{\Delta^{3/2}}, ~~E_z = \pm \frac{4b^2}
{\Delta^{3/2}}(t^2-z^2+\rho^2+b^2), ~~H_{\varphi} = \mp \frac{8b^2\rho t}
{\Delta^{3/2}},
\end{equation}
the remaining components being identically equal to zero.

For Born solution multiple interpretations were given, and the related
problem of radiation was discussed many times in literature. However, in almost all these works 
the second mirror charge was simply ignored \cite{Bol}-\cite{Ros}. In the last few years an 
increasing interest arose concerning the problem of radiation from a uniformly  accelerated charge 
\cite{Ng}-\cite{isr}, and highly qualified works appeared proving the absence of radiation in this case \cite{Sin1,Sin2}.  

Our point of view is similar to that of Singal's, though we have no 
opportunity to discuss here this old and confused problem. We note only that
the presence of a second charge seems to be an inevitable consequence
and a fundamental property of the Maxwell equations (see also \cite{isr1}).
The process itself (Fig.1) can be considered as a toy model for {\it elastic 
scattering} of two particle-like formations, interacting evidently in a non-electromagnetic 
way (repulsion instead of attraction!), and the charge plays the role of a field source and 
a conserved quantum number rather than a coupling constant~\footnote{Let us note, however, 
that the {\it effective} charge \cite{Str} $\sqrt{q^2 + (i \mu)^2}$, as for the unisingular 
solution (Compare with (\ref{eq18})) is identically equal to zero, formally justifying the absence of 
EM-interaction!}.

Such an interpretation can be confirmed by considering our {\it modifications}
of the Born solution corresponding to complex values of the constant 
$b^2=\alpha+i\beta,~\alpha,\beta \in \mathbb{R}$ in (\ref{eq19})\footnote{As before we deal here with solutions
of the ordinary {\it real} Maxwell equations.}.
In this case (supposing $\alpha,\beta \neq 0$) we have for singularities the
following relations:  
\begin{equation}\label{eq23}
\rho= \sqrt{\frac{\sqrt{\alpha^2+\beta^2}-\alpha}{2}},~~z=\pm \sqrt{t^2+\frac{\alpha+\sqrt{\alpha^2+\beta^2}}{2}}
\end{equation}

Hence, here again the singularities perform hyperbolic motion along $\pm Z$
axis but have a {\it ring-like structure} with fixed radia $\rho$ (\ref{eq23}).
In the same way as in the case of real constant $b$ we find again that
these ring-like formations {\bf have opposite charges equal in magnitude to the 
fundamental one.} 

The case of pure imaginary contant $b = ia, a\in \mathbb{R}$ requires special
consideration. From (\ref{eq20}) we have the following relation for caustic:
\begin{equation}\label{eq24}
z^2 + (\rho \pm a)^2 = t^2 .
\end{equation}
This, at first sight, simple equation describes a rather nontrivial evolution of
singularities. At $t=0$ we have a ring of radius $|a|$ lying in the $xy-$plane 
(Fig.2a), which at $t>0$ turns into a torus with a radius expanding at 
the speed of light (Fig.2b) (on these figures we represent a cross-section along the 
axis of symmetry $Z$). At $t=|a|$ the torus fills its gap (Fig.2c) and at the 
same time a second singularity given by the opposite sign in (\ref{eq24}) occurs.
\vskip5mm 
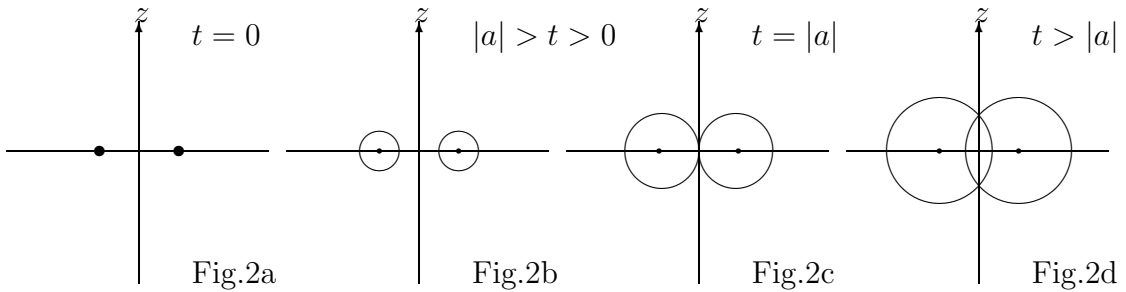
\begin{picture}(100,100)\\[10pt]
\put(50,0){\vector(0,1){99}}
\put(0,50){\line(1,0){99}}
\put(48,99){\jss{$z$}}
\put(70,90){\jss{$t=0$}}
\put(70,0){\jss{Fig.2a}}
\put(65,50){\circle*{4}}
\put(35,50){\circle*{4}}
\end{picture}
\begin{picture}(100,100)(-2,0)
\put(50,0){\vector(0,1){99}}
\put(0,50){\line(1,0){99}}
\put(48,99){\jss{$z$}}
\put(70,90){\jss{$|a|>t>0$}}
\put(70,0){\jss{Fig.2b}}
\put(65,50){\circle*{2}}
\put(35,50){\circle*{2}}
\put(65,50){\circle{15}}
\put(35,50){\circle{15}}
\end{picture}
\begin{picture}(100,100)(-4,0)
\put(50,0){\vector(0,1){99}}
\put(0,50){\line(1,0){99}}
\put(48,99){\jss{$z$}}
\put(70,90){\jss{$t=|a|$}}
\put(70,0){\jss{Fig.2c}}
\put(65,50){\circle*{2}}
\put(35,50){\circle*{2}}
\put(64,50){\circle{29}}
\put(36,50){\circle{29}}
\end{picture}
\begin{picture}(100,100)(-6,0)
\put(50,0){\vector(0,1){99}}
\put(0,50){\line(1,0){99}}
\put(48,99){\jss{$z$}}
\put(70,90){\jss{$t>|a|$}}
\put(70,0){\jss{Fig.2d}}
\put(65,50){\circle*{2}}
\put(35,50){\circle*{2}}
\put(65,50){\circle{40}}
\put(35,50){\circle{40}}
\end{picture}
\vskip5mm
At $t>|a|$ the torus keeps on expanding, penetrating through itself. After this 
the singularity has a form of two toroidal {\it "bridges"}, connecting their common
points with coordinates     
\begin{equation}\label{eq25}
\rho = 0, ~~~z =\pm \sqrt{t^2 - a^2}
\end{equation}
(the corresponding cross-section is represented on Fig.2d).

At $t\gg |a|$ the law of motion of these singularities approachs that of the
Born solution (\ref{eq21}). However, their velocity falls from $\infty$ to the
velocity of light. This makes the process to look like a run-away of two 
{\it tachion-like} formations. In the interval between $t=-\infty$ and $t=0$
we have an inverse process of accelerated rapprochement of these tachions
until their {\it annihilation} at $t=-|a|$, with formation of a toroidal 
{\it resonance} structure and its subsequent degeneration into a singular 
ring of radius $|a|$ at $t=0$. At this instant the electric lines configuration undergoes topological changes
since the winding number could be seen to change sign.  

Observe that asymptotically at $|t|\gg |a|$ the field of the tachion solution
coincides with that of Born solution for all directions determined by small
azimuths $\theta \ll {\pi}/2$ (i.e. far from directions to singular
bridges). This observation allows us to consider these tachion-like 
singularities (\ref{eq25}) as "quasi-charges", with charge magnitude equal again to that
of the fundamental solution. 

The metric correponding to this solution of GSE is of the Kerr-Schild type,
\begin{equation}\label{eq26}
g_{\mu \nu}=\eta_{\mu \nu}+he^3_\mu e^3_\nu
\end{equation}
where $e^3=du+Gdw+\bar{G}d\bar{w}+G\bar{G}dv$ is a null tetrad 1-form.\\
For real contants $b$ in analogy with~\cite{BKP,Kin} we have a real metric with the function 
\begin{equation}\label{eq27}
h=\frac{m\partial_u{\bar{G}}}{\bar{\tau}_2(\bar{\tau}_2G-\bar{\tau}_1)^2},~~m\equiv m(\bar{\tau}_1)
\end{equation}
satisfying the Einstein field equations for a null radiation matter tensor
$T_{\mu \nu}=~P_{33}e^3_\mu e^3_\nu$, where $P_{33}$ is given by

\begin{equation}\label{eq28}
P_{33}= 2\frac{(\dot{m}(\bar{\tau}_2G-\bar{\tau}_1)\bar{\tau}_1+3m(\bar{\tau}_1+G\bar{\tau}_2))(\bar{\tau}_2\partial_u\bar{G})^2}{(\bar{\tau}_2G-\bar{\tau}_1)^3},~~\dot{m}=\frac{dm}{d\bar{\tau}_1}.
\end{equation}
Note that the shear-free geodesic congruence defined by the function $G$ in (\ref{eq19}) has 
zero twist (i.e. pure expansion). Solution (\ref{eq19}) describes a curved space filled with null radiation.
The EM-field may be considered as a {\it test} field, since by the virtue of complex 
self-duality (\ref{eq4}) it does not contribute to the energy-momentum tensor.                     

\section{\hspace{-4mm}.\hspace{2mm}CONCLUSION. INTERACTIONS \& TRANSMUTATIONS OF PARTICLES IN  
BIQUATERNIONIC FIELD DYNAMICS}
\label{con}
The reduction of GSE (\ref{eq1}) to the algebraic solution of the shear-free geodesic congruence equation (\ref{eq13}) 
(or to equations (\ref{eq11}) in the arbitrary gauge) allows us in a fairly simple way to generate complex 
solutions of the GSE and corresponding to them solutions of the Maxwell
and ($\mathbb{C}$\,-valued) Yang-Mills equations. Until now other solutions among them 
an electrically neutral "photon-like" one with helix-like singularity structure have been studied 
(V.N. Trishin). A wide class of singular solutions of the free Maxwell equations
was given in~\cite{Bat}, however they still have to be adjusted with the basic GSE.

For more complicated distributions, their most interesting characteristic~-~the 
structure of singularity and its evolution in time~-~may be obtained without
explicitly solving equation (\ref{eq13}). In fact, excluding the main function $G(x)$
from (\ref{eq13}) and (\ref{eq14}) immediately results in the equation of singularity.
There is a variety of such complex singularities which will be published
elsewhere.

Nevertheless, the main two tasks naturally arising in our approach~-~the 
complete classification of singularities and derivation of their equations of
motion~-~remain unsolved, requiring probably more sophisticated mathematics. In view 
of the overdetermined GSE the Cauchy intial data can be specified for a 
{\it 2-dimensional surface} rather than in all 3-space, since evolution
in the third coordinate as well as in time are determined by system (\ref{eq1})
itself. 

As to the classification problem of singularities and the related problem of 
their admissible transmutations, it probably (at least for vacuum Maxwell equations) 
will be solved soon in the frame of general {\it catastrophe theory} 
\cite{Arn}. The identification of singularities with 
elementary particles automatically comes to mind. However physicists are not
used to such an approach, since there, singularities usually emerge as 
{\it divergences} of some integral characteristics (energy etc). But in the 
considered here non-Lagrangian approach this difficulty may be fictitious, 
since {\bf quantum numbers as well as the equations of motion for
these singularities are determined by solutions of field equations themselves.}  

Conceptually, we deal here with a new approach to nonlinear electrodynamics, where instead
of modifying Maxwell equations themselves we regard them as necessary consequences of
some overdetermined nonlinear predynamics.

Thus, behind a compact and following from a unique general principle (hyper-holomorphy of field functions) GSE
(\ref{eq1}) we have an entire world of structures close in many aspects to the real one. After appropriate interpretation
of mathematical variables arising, the language of the theory appears to be much similar (but not identical!) to
that used in customary field theory.

Moreover, some facts (such as quantization of electric charge), may be described in the framework of GSE (\ref{eq1})
much more naturally and adequately, than in conventional field-theoretical schemes. On the other hand, the true
potentialities of GSE (\ref{eq1}) in describing the structure and interaction of particles still have to be clarified.
Certainly, we don't regard GSE (\ref{eq1}) as an {\it ultimate} mathematical structure coding the 
"Theory of Everything". However, even its exclusive properties already obtained make this system a suitable and 
impressive demonstration of basic principles of algebrodynamical approach. 

Note that this scheme admits natural generalizations to algebras of higher dimensions as well as to 
{\it "local algebras"} with structure "constants" depending on space points~\cite[Chapter 2]{AD} and to nonassociative 
algebras like octonions. All these generalizations seem to be promising.

The authors are indebted to Yu.~S.~Vladimirov, Tz.~I.~Gutzunayev, D.~P.~Zhelobenko and Yu.~P.~Rybakov for useful advice and
interest exhibited. One of us (V.K.) is grateful to A.~V.~Aminova, D.~A.~Kalinin and other organizers of the school-seminar 
"Volga-10" for cordial reception, and to its participants, personally to V. G. Bagrov and A. P. Shirokov~-~for stimulating 
discussions.

\end{document}